\documentclass[preprint,12pt]{elsarticle}
\usepackage{epic}
\usepackage{eepic}
\usepackage{makeidx}
\usepackage{array}
\usepackage{times}
\usepackage{amssymb}
\usepackage{amsmath}
\usepackage{amsxtra}
\usepackage{graphicx}
\usepackage{subfig}
\usepackage{color}
\usepackage{boxedminipage}
\usepackage{alltt}
\usepackage{tikz}
\usepackage{booktabs}
\usepackage{xspace}
\usepackage[pdftex=true,hyperindex=true,colorlinks=true,breaklinks=true, urlcolor=red, linkcolor=blue, citecolor=black,bookmarksnumbered=true,pdfpagelabels=true,plainpages=false]{hyperref}
\usepackage{rotating}
\usepackage{float}
\usepackage{multirow}
\usepackage{listings}
\usepackage{url}

\usetikzlibrary{calc,fadings,decorations.markings}
\usetikzlibrary{positioning,shapes,shadows,arrows}
\usetikzlibrary{shapes.geometric}

\newcolumntype{C}{>{\centering\arraybackslash}m{2cm}}

\newcommand{\R}{{\mathbb R}}
\newcommand{\N}{{\mathbb N}}
\newcommand{\Z}{{\mathbb Z}}

\newlength{\mylablen}
\newlength{\mycaplen}
\newlength{\myislen}
\newcommand{\fcap}[2]{\refstepcounter{figure} \settowidth{\mycaplen}{#2}
\settowidth{\mylablen}{\bf Figure \thefigure:~}
\settowidth{\myislen}{\textwidth-\mylablen}
\begin{center}
{\bf Figure \thefigure:~}\ifthenelse{\mycaplen > \myislen}{\parbox[t]{\textwidth-\mylablen}{#2}\label{#1}}{#2\label{#1}}\end{center}}
\newcommand{\tcap}[2]{\refstepcounter{table} \settowidth{\mycaplen}{#2}
\settowidth{\mylablen}{\bf Table \thetable:~}
\settowidth{\myislen}{\textwidth-\mylablen}
\begin{center}
{\bf Table \thetable:~}\ifthenelse{\mycaplen > \myislen}{\parbox[t]{\textwidth-\mylablen}{#2}\label{#1}}{#2\label{#1}}\end{center}}
\newcommand{\ippl}{\textsc{IPPL}}
\newcommand{\classic}{\textsc{CLASSIC}}

\newcommand{\opal}{\textsc{OPAL}}

\def\C{{{{\rm {\mbox{\small l}}} \kern -.50em {\rm C}}}}
\def\R{{{{\rm l} \kern -.15em {\rm R}}}}
\def\N{{{{\rm l} \kern -.15em {\rm N}}}}
\def\E{{{{\rm l} \kern -.15em {\rm E}}}}
\def\P{{{{\rm l} \kern -.15em {\rm P}}}}
\def\D{{{{\rm l} \kern -.15em {\rm D}}}}
\def\L{{{{\rm l} \kern -.15em {\rm L}}}}
\def\Z{{{{\rm Z} \kern -.35em {\rm Z}}}}
\def\Q{{{{\rm {\mbox{\small l}}} \kern -.45em {\rm Q}}}}
\def\qQ{{{{\rm {\mbox{\scriptsize l}}} \kern -.33em {\rm Q}}}}

\def\-v{\vspace{-1ex}}

\def\i{\big\{}

\def\l{\ell^\prime}

\newcommand{\myvec}[1]{\bf{#1}}

\renewcommand{\vec}{\myvec}

\renewcommand{\epsilon}{\varepsilon}

\makeatletter
\pgfkeys{/pgf/decoration/.cd,
         start color/.store in = \startcolor,
         end color/.store in   = \endcolor
        }

\pgfdeclaredecoration{color change}{initial}{
\state{initial}[%
    width                     = 0pt,
    next state                = line,
    persistent precomputation = {\pgfmathdivide{50}{\pgfdecoratedpathlength}%
                                 \let\increment=\pgfmathresult%
                                 \def\x{0}}]%
{}%

\state{line}[%
    width                      = .5pt,
    persistent postcomputation = {\pgfmathadd@{\x}{\increment}%
                                  \let\x=\pgfmathresult}]%
{%
  \pgfsetlinewidth{\pgflinewidth}%
  \pgfsetarrows{-}%
  \pgfpathmoveto{\pgfpointorigin}%
  \pgfpathlineto{\pgfqpoint{.75pt}{0pt}}%
  \pgfsetstrokecolor{\endcolor!\x!\startcolor}%
  \pgfusepath{stroke}%
}%

\state{final}{%
  \pgfsetlinewidth{\pgflinewidth}%
  \pgfpathmoveto{\pgfpointorigin}%
  \color{\endcolor!\x!\startcolor}%
  \pgfusepath{stroke}%
}
}
\makeatother

\def\rnd{.%
\pdfuniformdeviate10%
\pdfuniformdeviate10%
\pdfuniformdeviate10%
}

\journal{Computer Physics Communications}

\begin{document}

\definecolor{sand}{RGB}{193,154,107} 
\def\cols{20}                        
\def\rows{40}                        
\def\SquareUnit{.35}                 
\pgfmathsetmacro\RmaxParticle{.1}    
\def\BeforeLight{5}                  

\begin{frontmatter}



\title{The Dynamical Kernel Scheduler - Part 1}


\author[label1]{Andreas Adelmann\corref{mycorrespondingauthor}}
\cortext[mycorrespondingauthor]{Corresponding author}
\ead{andreas.adelmann@psi.ch}

\author[label1,label3]{Uldis Locans}

\author[label1]{Andreas Suter}
\address[label1]{Paul Scherrer Institut, Villigen, CH-5232, Switzerland}
\address[label3]{University of Latvia, 19 Raina Blvd., Riga, LV 1586, Latvia}

\begin{abstract}

Emerging processor architectures such as GPUs and Intel MICs provide a huge performance potential for high performance computing. However developing software that uses these hardware accelerators introduces additional challenges for the developer. These challenges may include exposing increased parallelism, handling different hardware designs, and using multiple development frameworks in order to utilise devices from different vendors.

The Dynamic Kernel Scheduler (DKS) is being developed in order to provide a software layer between the host application and different hardware accelerators. DKS handles the communication between the host and the device, schedules task execution, and provides a library of built-in algorithms. Algorithms available in the DKS library will be written in CUDA, OpenCL, and OpenMP. Depending on the available hardware, the DKS can select the appropriate implementation of the algorithm.

The first DKS version was created using CUDA for the Nvidia GPUs and OpenMP for Intel MIC. DKS was further integrated into OPAL (Object-oriented Parallel Accelerator Library) in order to speed up a parallel FFT based Poisson solver and Monte Carlo simulations for particle matter interaction used for proton therapy degrader modelling. DKS was also used together with Minuit2 for parameter fitting, where $\chi^2$ and max-log-likelihood functions were offloaded to the hardware accelerator. The concepts of the DKS, first results, and plans for the future will be shown in this paper.
\end{abstract}


\end{frontmatter}


\section{Introduction}
\label{sec:introduction}

In recent years hardware accelerators have become increasingly popular within scientific computing. Based on the Top500 list from June 2015 \cite{top500}, 90 of the top 500 supercomputers in the world are accelerator based. This includes the top two systems on the list: Tianhe-2 which uses Intel Xeon Phi coprocessors and Titan which uses NVIDIA K20x GPUs. GPU usage for general purpose computing has become even more important, due to the gaming industry. Almost every computer is now equipped with a GPU, but if the application is not exploiting the GPU, it is not using all the available computational power of the system. However, developing software that can take advantage of hardware accelerators can become a challenging task, especially for large existing applications. Each hardware accelerator has its own architecture and memory hierarchy, which must be taken into account to gain the maximum performance out of the device. In addition to hardware differences, there are also varying methods to program these devices. NVIDIA provides the CUDA \cite{cuda} toolkit for its GPUs, both AMD and NVIDIA support the OpenCL \cite{opencl} framework, and Intel allows usage of standard tools and languages to program Intel MIC processor \cite{mic}. There are also OpenACC \cite{openacc} and OpenMP \cite{openmp} standards that allow the targeting of hardware accelerators by expressing parallelism through compiler directives.

In this work, the Dynamic Kernel Scheduler (DKS) is presented which provides a slim software layer between the host application and the hardware accelerators. DKS separates the accelerator and framework specific code from the host application and provides a simple interface that can be implemented in the host application to offload tasks to the accelerator. DKS provides functions to handle communication and data transfer between host and device, as well as a library of functions written in CUDA, OpenCL, and OpenMP that allow the targeting of different accelerators. The first version of DKS was integrated into OPAL (Object-oriented Parallel Accelerator Library). This DKS version uses CUDA kernels and OpenMP offload pragmas to run OPAL's FFT based Poisson solver and Monte Carlo simulations on a GPU and Intel MIC. DKS was also used together with Minuit2 for parameter fitting, where $\chi^2$ and max-log-likelihood functions were offloaded to the hardware accelerator.

In the literature, there are several FFT Based Poisson solvers developed for GPUs using CUDA which use NVIDIA's cuFFT library \cite{Decyk2011,Rossinelli2010}. One can also find research on the use of customized FFTs for asynchronous execution and mapping FFT based Poisson solvers to multi node systems \cite{Dugan2013,Wu2013,Wu2014}. Numerous studies \cite{Osiecki2013,Jia2012,Jia2011,Liu2015,Xu2015,Tickner2010} have been carried out to show the potential of GPUs and Intel Xeon Phi co-processors for Monte-Carlo simulations for proton and photon transport. These problems are some of the most time consuming parts of the OPAL simulations, and previous research shows that they are good candidates for acceleration on the co-processors.

Many research projects try to focus on improving programmability of hardware accelerators. Apart from compiler directive based approaches, there are a number of vendor supported libraries \cite{nvidialibraries,miclibraries} that allow the simplification of offloading specific tasks to accelerators. There has also been work on creating abstractions and providing software layers that would allow to express kernels \cite{Bourgoin2014,Svensson2010,Vinas2015} for hardware accelerators, which can be translated to CUDA or OpenCL code that is run on the device.

The ability of DKS to have implementations using different frameworks and libraries, and switch between them from the host application allows the targeting of hardware accelerators of different types and fine tuning of the code to gain the maximum performance from each device. This approach also provides moro portability and software investment protection for the host application. In case some hardware architecture is no longer manufactured, or some new architecture or development framework emerges, only DKS needs to be updated.

The rest of the paper is structured as follows: Section \ref{sec:concarch} describes the concepts and architecture of DKS; Section \ref{sec:dksappl} describes the concepts of OPAL's FFT based Poisson solver and Monte-Carlo type particle matter interaction simulations as well as DKS implementation of these functions and the benchmark results; Section \ref{sec:paramfit} explains the DKS and \textsc{Minuit2} usage and results for parameter fitting using hardware acceleration; and Section \ref{sec:concloutl} provides conclusions and future of the DKS.

\section{Concept and Architecture of the Dynamic Kernel Scheduler (DKS)}
\label{sec:concarch}

\subsection{Concept}
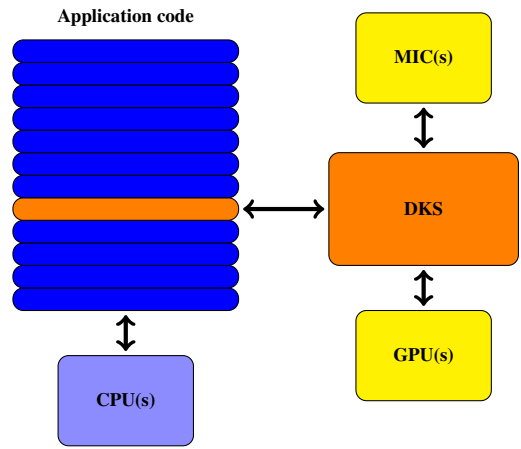
\begin{figure}[htb]
  \centering
  \begin{tikzpicture}[scale=0.6, transform shape]
    \begin{scope}[shape=rectangle,rounded corners,text centered]
 
      \draw [fill=blue!45] (1,0) rectangle (4,2);
      \node at (2.5,1) {\textbf{CPU(s)}};

      \draw [ultra thick] [<->] (2.5, 2.1) -- (2.5, 2.9);

      \draw [fill=blue] (0,3) rectangle (5,3.5);
      \draw [fill=blue] (0,3.5) rectangle (5,4);
      \draw [fill=blue] (0,4) rectangle (5,4.5);
      \draw [fill=blue] (0,4.5) rectangle (5,5);
      \draw [fill=orange] (0,5) rectangle (5,5.5);
      \draw [fill=blue] (0,5.5) rectangle (5,6);
      \draw [fill=blue] (0,6) rectangle (5,6.5);
      \draw [fill=blue] (0,6.5) rectangle (5,7);
      \draw [fill=blue] (0,7) rectangle (5,7.5);
      \draw [fill=blue] (0,7.5) rectangle (5,8);
      \draw [fill=blue] (0,8) rectangle (5,8.5);
      \draw [fill=blue] (0,8.5) rectangle (5,9);
      \node at (2.5, 9.5) {\textbf{Application code}};

      \draw [ultra thick] [<->] (5.1,5.25) -- (6.9, 5.25);
      \draw [fill=orange] (7,4) rectangle (11.2,6.5);
      \node at (9.1, 5.25) {\textbf{DKS}};

      \draw [ultra thick] [<->] (9.1, 3.9) -- (9.1, 3.1);
      \draw [ultra thick] [<->] (9.1, 6.6) -- (9.1, 7.5);

      \draw [fill=yellow] (7.6, 1) rectangle (10.6, 3);
      \node at (9.1, 2) {\textbf{GPU(s)}};
      
      \draw [fill=yellow] (7.6, 7.6) rectangle (10.6, 9.6);
      \node at (9.1, 8.6) {\textbf{MIC(s)}};
    
    \end{scope}
  \end{tikzpicture}
  \caption{The Dynamic Kernel Scheduler concept}
  \label{fig:dksconcept}
\end{figure}

The Dynamic Kernel Scheduler (DKS) is a slim software layer between the host application and the hardware accelerator, as depicted in Figure ~\ref{fig:dksconcept}. The aim of the DKS is to allow the creation of fast fine tuned kernels using device specific frameworks such as CUDA, OpenCL, OpenACC and OpenMP. On top of that, DKS allows the easy use of these kernels in host applications without providing any device or framework specific details. This approach facilitates the integration of different types of devices in the existing applications with minimal code changes and  makes the device and the host code a lot more manageable.

The architecture of DKS can be split in three main parts: 
\begin{enumerate}
\item The first part provides communication functions that handle memory allocation and data transfer to, and from, the device. All the memory management is left up to the user. So that the data transfers and memory allocation can be scheduled only when necessary. DKS also supports GPU streams such that asynchronous data transfer and kernel execution can be implemented when possible. 
\item The second part of DKS consists of a function library, which contains algorithms written in CUDA, OpenCL, and OpenMP to target different devices. DKS can switch between implementations based on the hardware that is available. Writing functions using multiple frameworks results in extra work, but this provides the opportunity to fine tune kernels for each device architecture for maximum performance. That also allows the targeting of systems containing different types of devices. The different implementations of the code are always separated so the code is still easy to manage. Additionally if a host application is targeted at a specific system, implementations that are not needed can be omitted. 
\item The third part of DKS is the auto-tuning functionality which will be discussed in a forthcoming paper. The aim of auto-tuning is to select the appropriate implementation of the algorithm and change the launch parameters according to the devices that are available on the system in order to gain the maximum performance. The auto-tuning functionality relies on knowledge of device architecture and benchmark tests that can be run on the system before running the application.
\end{enumerate}

\begin{figure}[htb]
  \centering
  \lstset{language=C++,
    basicstyle=\ttfamily\footnotesize,
    keywordstyle=\color{blue}\ttfamily\footnotesize,
    commentstyle=\color{red}\ttfamily\footnotesize,
    breaklines=true}
  \begin{lstlisting}[frame=single]
    //allocate memory on device and write data
    void *mem_ptr;
    mem_ptr = dks.allocateMemory<Complex_t>(DATA_SIZE, NULL);
    dks.writeData<Complex_t>(mem_ptr, DATA_ARRAY, DATA_SIZE);

    //execute FFT or IFFT
    if (direction == 1)
      dks.callFFT(mem_ptr, DIMENSIONS, DIM_SIZE);
    else
      dks.callIFFT(mem_ptr, DIMENSIONS, DIM_SIZE);

    //read data and free memory
    dks.readData<Complex_t>(mem_ptr, DATA_ARRAY, DATA_SIZE);
    dks.freeMemory<Complex_t>(mem_ptr, DATA_SIZE);
  \end{lstlisting}
  \caption{Example of DKS usage for FFT}
  \label{fig:dksexample}
\end{figure}

Figure \ref{fig:dksexample} shows an example code of DKS usage inside a host application to perform Fast Fourier transform. The host application has full control over the memory allocation and data transfer to the device, but there are no device specific details in the host code. DKS evaluates the calls made by host application and chooses the appropriate device to use, and algorithm implementation, to run the code on selected accelerator.

\subsection{Architecture}
\begin{figure}[htb]
  \centering
  \tikzstyle{abstract}=[rectangle, draw=black, rounded corners, fill=blue!30,
        text centered, anchor=north, text=black, text width=3cm]
\tikzstyle{notdone}=[rectangle, draw=black, rounded corners, fill=red!30,
        text centered, anchor=north, text=black, text width=3cm]
\tikzstyle{comment}=[text height=2cm]
\tikzstyle{myarrow}=[->, >=open triangle 90, thick]
\tikzstyle{line}=[-, thick]

\begin{tikzpicture}[scale=0.5, transform shape]

\node (DKSBase) [abstract, rectangle split, rectangle split parts=2]
      {
        \textbf{DKSBase}
        \nodepart{second}Handle calls from host application
      };

\node (DKSAux) [comment, text width=4cm, left=of DKSBase] {};

\node (AutoTune) [notdone, rectangle split, rectangle split parts=2, left=of DKSAux]
      {
        \textbf{DKSAutoTuning}
        \nodepart{second}Handle autotuning of the DKS library
      };

\node (OpenCLBase) [abstract, rectangle split, rectangle split parts=2, below=of DKSBase]
      {
        \textbf{OpenCLBase}
        \nodepart{second}Handle OpenCL device, memory, data transfer
      };

\node (Left1) [text width=4cm, left=of OpenCLBase] {};
\node (Right1) [text width=4cm, right=of OpenCLBase] {};

\node (CUDABase) [abstract, rectangle split, rectangle split parts=2, left=of Left1]
      {
        \textbf{CUDABase}
        \nodepart{second}Handle CUDA device, memory, data transfer
      };

\node (MICBase) [abstract, rectangle split, rectangle split parts=2, right=of Right1]
      {
        \textbf{MICBase}
        \nodepart{second}Handle MIC device, memory, data transfer
      };

\node (CUDAAux) [comment, below=of CUDABase] {};

\node (CUDACollimator) [abstract, text width=4.5cm, rectangle split, rectangle split parts=2, below=of CUDAAux]
      {
        \textbf{CUDA CollimatorPhysics}
        \nodepart{second}Functions for OPALs Collimator physics solver offloads
      };

\node (CUDAFFT) [abstract, rectangle split, rectangle split parts=2, left=of CUDAAux]
      {
        \textbf{CUDA FFTPoisson}
        \nodepart{second}Functions for OPALs FFT solver offloads
      };

\node (CUDAChi) [abstract, rectangle split, rectangle split parts=2, right=of CUDAAux]
      {
        \textbf{CUDA ChiSquare}
        \nodepart{second}Functions for Chi-square and max-log-likelihood function offload
      };

\node (OpenCLAux) [comment, below=of OpenCLBase] {};

\node (OpenCLCollimator) [notdone, text width=4.5cm, rectangle split, rectangle split parts=2, below=of OpenCLAux]
      {
        \textbf{OpenCL CollimatorPhysics}
        \nodepart{second}Functions for OPALs Collimator physics solver offloads
      };

\node (OpenCLFFT) [notdone, rectangle split, rectangle split parts=2, left=of OpenCLAux]
      {
        \textbf{OpenCL FFTPoisson}
        \nodepart{second}Functions for OPALs FFT solver offloads
      };

\node (OpenCLChi) [abstract, rectangle split, rectangle split parts=2, right=of OpenCLAux]
      {
        \textbf{OpenCL ChiSquare}
        \nodepart{second}Functions for Chi-square and max-log-likelihood function offload
      };

\node (MICAux) [comment, below=of MICBase] {};

\node (MICCollimator) [abstract, text width=4.5cm, rectangle split, rectangle split parts=2, below=of MICAux]
      {
        \textbf{MIC CollimatorPhysics}
        \nodepart{second}Functions for OPALs Collimator physics solver offloads
      };

\node (MICFFT) [abstract, rectangle split, rectangle split parts=2, left=of MICAux]
      {
        \textbf{MIC FFTPoisson}
        \nodepart{second}Functions for OPALs FFT solver offloads
      };

\node (MICChi) [abstract, rectangle split, rectangle split parts=2, right=of MICAux]
      {
        \textbf{MIC ChiSquare}
        \nodepart{second}Functions for Chi-square and max-log-likelihood function offload
      };

\draw[myarrow] (CUDABase.north) -- ++(0,0.5) -| (DKSBase.south);
\draw[myarrow] (OpenCLBase.north) -- ++(0,0.5) -| (DKSBase.south);
\draw[myarrow] (MICBase.north) -- ++(0,0.5) -| (DKSBase.south);

\draw[myarrow] (CUDAFFT.east) -- ++(0,0.1) -| (CUDABase.south);
\draw[myarrow] (CUDAChi.west) -- ++(0,0.1) -| (CUDABase.south);
\draw[myarrow] (CUDACollimator.north) -- (CUDABase.south);

\draw[myarrow] (OpenCLFFT.east) -- ++(0,0.1) -| (OpenCLBase.south);
\draw[myarrow] (OpenCLChi.west) -- ++(0,0.1) -| (OpenCLBase.south);
\draw[myarrow] (OpenCLCollimator.north) -- (OpenCLBase.south);

\draw[myarrow] (MICFFT.east) -- ++(0,0.1) -| (MICBase.south);
\draw[myarrow] (MICChi.west) -- ++(0,0.1) -| (MICBase.south);
\draw[myarrow] (MICCollimator.north) -- (MICBase.south);

\draw[myarrow] (AutoTune.east) -- (DKSBase.west);

\end{tikzpicture}
  \caption{Architecture of the Dynamic Kernel Scheduler. The OpenCL implementation and the auto tuning framework are both shown in red and will be discussed in a subsequent publication.}
  \label{fig:dksarch}
\end{figure}
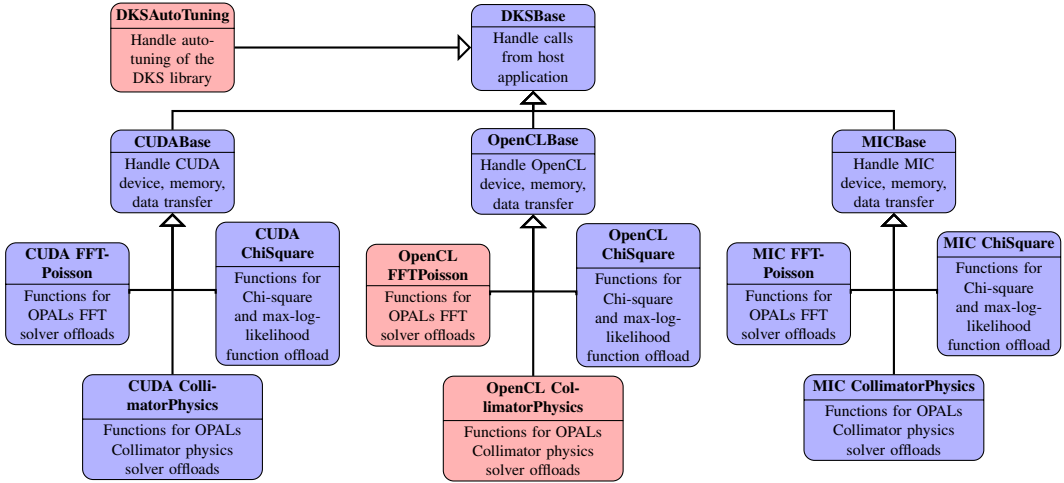

The Dynamic Kernel Scheduler is split into separate modules. Each module contains function implementations using different frameworks. The base class for each module contains functions which handles the device management, memory management, and data transfer, this base class can be extended to cover all the necessary algorithm specific functions. The base class of DKS receives all the calls from the host application and decides which device specific implementation should be used to run the code on the device. Figure ~\ref{fig:dksarch} shows the architecture of the first version of DKS, for each module base class can be easily extended to include other algorithms and the base class of DKS can be extended to include other modules to handle different development frameworks.

\section{The Dynamic Kernel Scheduler used in Object-oriented Parallel Accelerator Library (OPAL)}
\label{sec:dksappl}

\label{subsec:dksopal}
\opal\ (Object Oriented Particle Accelerator Library) is a parallel, open source C++ framework for general particle accelerator simulations which includes 3D space charge, short range wake fields, and particle matter interaction. \opal\ is based on \ippl\ (Independent Parallel Particle Layer) which adds parallel capabilities. Main functions inherited from \ippl\ are structured rectangular grids, fields, parallel FFT, and particles with the respective interpolation operators. Other features are expression templates, and massive parallelism (up to 65000 processors) which allows it to tackle the largest problems in the field. 
\begin{figure}[htb]
\begin{center}
 \begin{tikzpicture}[scale=0.73, transform shape]
    \footnotesize
      \begin{scope}[shape=rectangle,rounded corners,minimum width=3.0cm,minimum height=0.6cm,fill=yellow,text centered]
       \node[thick, anchor=center, opacity=1., font=\Large] at (3.8,2) {\textbf{\opal}};
       \node[fill= green!45] (0_00) at (0.1,1.0) {MAD-Parser};
       \node[fill= green!40] (0_00) at (3.5,1.0) {Flavors: t,Cycl,Map};
       \node[fill= green!40] (0_00) at (7.0,1.0) {Distributions};
       \node[fill= green!40] (0_00) at (0.1,0.0) {Solvers: Direct,MG};
       \node[fill= green!40] (0_00) at (3.5,0.0) {Integrators};
       \node[fill= green!40] (0_00) at (7.0,0.0) {Particle Matter Inter.};
       \node[fill= red!45] (q_00) at (0.1,-1) {FFT};
       \node[fill= red!45] (q_01) at (3.5,-1) {D-Operators};
       \node[fill= red!45] (q_02) at (7,-1) {NGP,CIC,TSI};
       \node[fill= red!45] (q_10) at (0.1,-1.75) {Fields};
       \node[fill= red!45] (q_11) at (3.5,-1.75) {Mesh};
       \node[fill= red!45] (q_12) at (7,-1.75) {Particles};
       \node[fill=red!45] (q_20) at (0.1,-2.5) {Load Balancing};
       \node[fill=red!45] (q_21) at (3.5,-2.5) {Domain Decomp.};
       \node[fill=red!45] (q_22) at (7,-2.5) {Communication};
       \node[fill=red!45] (q_20) at (0.1,-3.25) {Particle-Cache};
       \node[fill=red!45] (q_21) at (3.5,-3.25) {PETE};
       \node[fill=red!45] (q_22) at (7,-3.25) {Trillions Interface};
       \node[rotate=90,minimum width=1.7cm,fill=gray] (bla) at (-1.9,0.49) {\classic};
       \node[rotate=90,minimum width=1.5cm,fill= magenta] (bla) at (-1.9,-1.38) {H5hut};
       \node[rotate=90,minimum width=1.4cm,fill= yellow] (bla) at (-1.9,-2.88) {BOOST};
       \node[fill=blue!65,minimum width=10.75cm] (q_23) at (3.1,-4.0) {Trilinos \& GSL};

       \draw[<->, color=orange!75, fill=orange!75, line width=3pt] (8.6, -0.55) to (10.1, -0.55);
       \node[thick, anchor=center, opacity=1., font=\Large] at (13.3, 2) {\textbf{DKS}};
       \node[rotate=90,minimum width=3.4cm,fill=orange] (bla) at (10.5,-0.55) {DKS API};

       \draw[rounded corners, draw=orange!25, thick, fill=orange!25, opacity=0.5, text centered] (10.9, 1.1) rectangle (16.1,-0.5) node[pos=0.5, black, thick, anchor=center, opacity=1., font=\small] {CUDA};
       \draw[rounded corners, draw=orange!45, thick, fill=orange!45, opacity=0.5, text centered] (11, 1) rectangle (13.4,0.5) node[pos=0.5, black, thick, anchor=center, opacity=1., font=\small] {cuFFT};
       \draw[rounded corners, draw=orange!45, thick, fill=orange!45, opacity=0.5, text centered] (13.6,1) rectangle (16,0.5) node[pos=0.5, black, thick, anchor=center, opacity=1., font=\small] {cuBLAS};
       \draw[rounded corners, draw=orange!45, thick, fill=orange!45, opacity=0.5, text centered] (11, 0.1) rectangle (13.4,-0.4) node[pos=0.5, black, thick, anchor=center, opacity=1., font=\small] {cuFFT};
       \draw[rounded corners, draw=orange!45, thick, fill=orange!45, opacity=0.5, text centered] (13.6, 0.1) rectangle (16,-0.4) node[pos=0.5, black, thick, anchor=center, opacity=1., font=\small] {cuBLAS};

       \draw[rounded corners, draw=orange!25, thick, fill=orange!25, opacity=0.5, text centered] (10.9, -0.6) rectangle (16.1,-2.2) node[pos=0.5, black, thick, anchor=center, opacity=1., font=\small] {MIC};       
       \draw[rounded corners, draw=orange!45, thick, fill=orange!45, opacity=0.5, text centered] (11, -0.7) rectangle (13.4,-1.2) node[pos=0.5, black, thick, anchor=center, opacity=1., font=\small] {OpenMP};
       \draw[rounded corners, draw=orange!45, thick, fill=orange!45, opacity=0.5, text centered] (13.6, -0.7) rectangle (16,-1.2) node[pos=0.5, black, thick, anchor=center, opacity=1., font=\small] {Offload};
       \draw[rounded corners, draw=orange!45, thick, fill=orange!45, opacity=0.5, text centered] (11, -1.6) rectangle (13.4,-2.1) node[pos=0.5, black, thick, anchor=center, opacity=1., font=\small] {Intel MKL};
       \draw[rounded corners, draw=orange!45, thick, fill=orange!45, opacity=0.5, text centered] (13.6, -1.6) rectangle (16,-2.1) node[pos=0.5, black, thick, anchor=center, opacity=1., font=\small] {Intel TBB};
      \end{scope}
 \end{tikzpicture}
\caption{The \opal\ software structure and connection to DKS}
\label{fig:opalstr}
\end{center}
\end{figure}
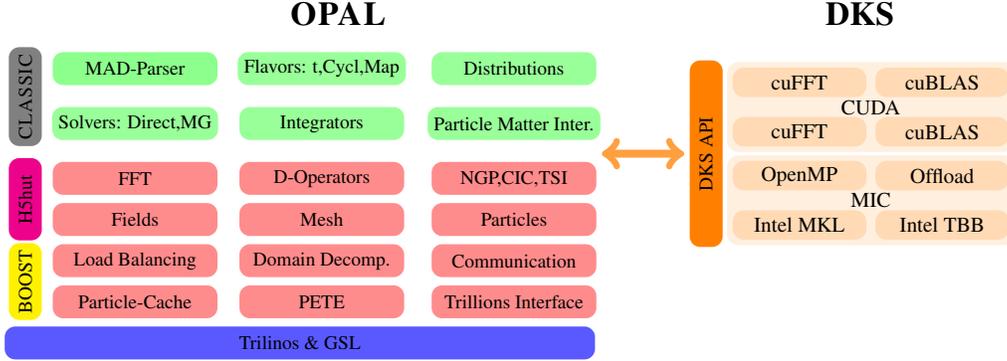

\subsection{FFT Based Particle-Mesh Solver}

The Particle-Mesh (PM) solver solver is best described in the book by R.W.~Hockney \& J.W.~Eastwood \cite{hockney1970methods}.
Instead of calculation the mutual interaction of a large number of particles, in the PM solver one discretises the computational domain
$\Omega:=[-a_x,a_x]\times[-a_y,a_y]\times[-a_z,a_z]$  into a regular mesh of $M_x\times M_y\times M_z$ grid points. The beam sizes 
$a_{x},a_{y},a_{z}$ are usually time dependent. 
Other geometries are possible but not discussed here. The mesh sizes $h_x,h_y$, and $h_z$ are allowed to change independently over time to assure a particle fitted grid. An essential part of  any self-consistent electrostatic beam dynamics code is the Poisson solver. From the -- time to solution -- point of view, we observe that in the order of $1/3$ of the computational time is spend in this algorithm,

In many of the physics application, the bunch can be considered as small compared to the transverse size of the surrounding beam pipe ($\partial \Omega$).\  If this is the case the conducting walls can be neglected and, we can solve an open boundary problem.\ Here we follow the method of Hockney  and compute the potential on a grid of size $2^3 M_x  M_y M_z$, assuming 3 spatial dimensions of the problem. The calculation then is making use of Fast Fourier Transform (FFT) techniques, with a computational effort scaling as 
$\mathcal{O}(2^3 M_x  M_y M_z (\log_2 2 M_x  M_y M_z)^3)$ \cite{hockney1970methods, eastwoodandbrownrigg,Hockney}.

\subsubsection{FFT-based Convolutions}
Given a charge density $\rho$, we search or the scalar potential $\phi$ by solving Poisson's equation
\begin{equation*}
\nabla^2\phi=-\rho/\epsilon_0,
\end{equation*} 
subject to $\phi = 0$ at $\partial \phi \rightarrow \infty$, i.e.\ in an unbound domain. If we know the Green's function $G(x,x',y,y',z,z')$, then the solution \begin{equation*}
\phi(x,y,z)=\int\int\int{dx' dy' dz'}\rho(x',y',z') G(x,x',y,y',z,z')
\end{equation*}
is the convolution of the a source charge at $(x',y',z')$ and $G$.\
In our case of an isolated charge distribution, we get
\begin{equation}
\phi(x,y,z)=\int\int\int{dx' dy' dz'}\rho(x',y',z') G(x-x',y-y',z-z'),
\label{eq:convolutionsol}
\end{equation}
with
\begin{equation*}
G(u,v,w)={1\over \sqrt{u^2+v^2+w^2}}.
\end{equation*}
We now discretise Eq.~(\ref{eq:convolutionsol})
on the previous mentioned Cartesian grid 
\begin{equation}
\phi_{i,j,k}=h_x h_y h_z \sum_{i'=1}^{M_x}\sum_{j'=1}^{M_y}\sum_{k'=1}^{M_z}  \rho_{i',j',k'}G_{i-i',j-j',k-k'}.
\label{eq:openbruteforceconv}
\end{equation}
The two scalar fields $\rho_{i,j,k}$ and $G_{i-i',j-j',k-k'}$ are now defined on the grid and we efficiently obtain the 
solution of Eq.~(\ref{eq:openbruteforceconv}) using Fourier techniques by
\begin{equation*}
\phi_{i,j,k}=h_x h_y h_z \text{ FT}^{-1} \{ ( \text{FT}\{\rho_{i,j,k}\}) \otimes ( \text{FT}\{G_{i,j,k}\}) \},
\label{oneterm}
\end{equation*}
with $\otimes$ denoting the Hadamard product.\ The notation $\text{FT}\{ . \}$ for the forward FFT and $\text{FT}^{-1}\{ . \}$ for the inverse FFT is used.

\subsubsection{The DKS Implementation of the Poisson Solver}
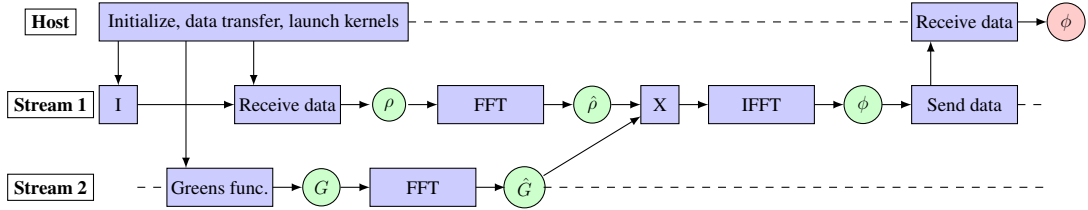
\begin{figure}[htb]
  \centering
\begin{tikzpicture}[scale=0.6, every node/.style={transform shape, font=\normalsize, minimum height=0.5cm,minimum width=0.5cm}]

\def\CPU{Host}
\def\GPU{Stream 1}
\def\GGPU{Stream 2}

\node [matrix, very thin,column sep=0.4cm,row sep=0.3cm] (matrix) at (0,0) {
  & \node(0,0) (\CPU) {}; & \node(0,0) (\CPU 0) {}; & \node(0,0) (\CPU 1) {}; & 
  \node(0,0) (\CPU 2) {}; & \node(0,0) (\CPU 3) {}; & \node(0,0) (\CPU 4) {}; & 
  \node(0,0) (\CPU 5) {}; & \node(0,0) (\CPU 6) {}; & \node(0,0) (\CPU 7) {}; & 
  \node(0,0) (\CPU 8) {}; & \node(0,0) (\CPU 9) {}; & \node(0,0) (\CPU 10) {}; & 
  \node(0,0) (\CPU 11) {}; & \node(0,0) (\CPU 12) {}; & \node(0,0) (\CPU 13) {}; &
  \node(0,0) (\CPU 14) {}; & \node(0,0) (\CPU 15) {}; & \node(0,0) (\CPU 16) {}; &\\
  & & & & & & & & & & & & & & & & & &\\ 
  & \node(0,0) (\GPU) {}; & \node(0,0) (\GPU 0) {}; & \node(0,0) (\GPU 1) {}; & 
  \node(0,0) (\GPU 2) {}; & \node(0,0) (\GPU 3) {}; & \node(0,0) (\GPU 4) {}; & 
  \node(0,0) (\GPU 5) {}; & \node(0,0) (\GPU 6) {}; & \node(0,0) (\GPU 7) {}; & 
  \node(0,0) (\GPU 8) {}; & \node(0,0) (\GPU 9) {}; & \node(0,0) (\GPU 10) {}; & 
  \node(0,0) (\GPU 11) {}; & \node(0,0) (\GPU 12) {}; & \node(0,0) (\GPU 13) {}; &
  \node(0,0) (\GPU 14) {}; & \node(0,0) (\GPU 15) {}; & \node(0,0) (\GPU 16) {};\\
  & & & & & & & & & & & & & & & & & &\\
  & \node(0,0) (\GGPU) {}; & \node(0,0) (\GGPU 0) {}; & \node(0,0) (\GGPU 1) {}; & 
  \node(0,0) (\GGPU 2) {}; & \node(0,0) (\GGPU 3) {}; & \node(0,0) (\GGPU 4) {}; & 
  \node(0,0) (\GGPU 5) {}; & \node(0,0) (\GGPU 6) {}; & \node(0,0) (\GGPU 7) {}; & 
  \node(0,0) (\GGPU 8) {}; & \node(0,0) (\GGPU 9) {}; & \node(0,0) (\GGPU 10) {}; & 
  \node(0,0) (\GGPU 11) {}; & \node(0,0) (\GGPU 12) {}; & \node(0,0) (\GGPU 13) {}; &
  \node(0,0) (\GGPU 14) {}; & \node(0,0) (\GGPU 15) {}; & \node(0,0) (\GGPU 16) {};\\
};

\fill 
(\CPU) node[draw,fill=white] {\textbf{\CPU}}
(\GPU) node[draw,fill=white] {\textbf{\GPU}}
(\GGPU) node[draw,fill=white] {\textbf{\GGPU}};

\filldraw[fill=blue!20]
(\CPU 0.north west) rectangle (\CPU 4.south east) node[pos=.5] {Initialize, data transfer, launch kernels}
(\GPU 0.north west) rectangle (\GPU 0.south east) node[pos=.5] {I}
(\CPU 12.north west) rectangle (\CPU 13.south east) node[pos=.5] {Receive data}
(\GPU 2.north west) rectangle (\GPU 3.south east) node[pos=.5] {Receive data}
(\GPU 5.north west) rectangle (\GPU 6.south east) node[pos=.5] {FFT}
(\GPU 8.north west) rectangle (\GPU 8.south east) node[pos=.5] {X}
(\GPU 9.north west) rectangle (\GPU 10.south east) node[pos=.5] {IFFT}
(\GPU 12.north west) rectangle (\GPU 13.south east) node[pos=.5] {Send data}
(\GGPU 1.north west) rectangle (\GGPU 2.south east) node[pos=.5] {Greens func.}
(\GGPU 4.north west) rectangle (\GGPU 5.south east) node[pos=.5] {FFT};

\draw
(\CPU 14) node[draw,circle,fill=red!20] {$\phi$}
(\GPU 4) node[draw,circle,fill=green!20] {$\rho$}
(\GPU 7) node[draw,circle,fill=green!20] {$\hat{\rho}$}
(\GPU 11) node[draw,circle,fill=green!20] {$\phi$}
(\GGPU 3) node[draw,circle,fill=green!20] {$G$}
(\GGPU 6) node[draw,circle,fill=green!20] {$\hat{G}$};


\draw [dashed] (\CPU 4) -- (\CPU 12);
\draw [-latex] (\CPU 13) -- (\CPU 14);

\draw [-latex] (\GPU 0) -- (\GPU 2);
\draw [-latex] (\GPU 3) -- (\GPU 4);
\draw [-latex] (\GPU 4) -- (\GPU 5);
\draw [-latex] (\GPU 6) -- (\GPU 7);
\draw [-latex] (\GPU 7) -- (\GPU 8);
\draw [-latex] (\GPU 8) -- (\GPU 9);
\draw [-latex] (\GPU 10) -- (\GPU 11);
\draw [-latex] (\GPU 11) -- (\GPU 12);
\draw [dashed] (\GPU 13) -- (\GPU 14);

\draw [dashed] (\GGPU 0) -- (\GGPU 1);
\draw [-latex] (\GGPU 2) -- (\GGPU 3);
\draw [-latex] (\GGPU 3) -- (\GGPU 4);
\draw [-latex] (\GGPU 5) -- (\GGPU 6);
\draw [-latex] (\GGPU 6) -- (\GPU 8);
\draw [dashed] (\GGPU 6) -- (\GGPU 14);

\draw [-latex] (\CPU 0) -- (\GPU 0);
\draw [-latex] (\CPU 1) -- (\GGPU 1);
\draw [-latex] (\CPU 2) -- (\GPU 2);
\draw [-latex] (\GPU 12) -- (\CPU 12);

\end{tikzpicture}
  \caption{FFT-Poisson solver sequence diagram}
  \label{fig:fft_sd}
\end{figure}

For use on NVIDIA GPUs the FFT Poisson solver is implemented in DKS using CUDA. It uses NVIDIA's cuFFT library to perform the FFT, separate kernels to calculate the Greens function
and, perform the multiplication on the GPU. CUDA streams are used to overlap the transfer of the $\rho$
field to the GPU and the calculation of the Greens function. The sequence diagram in Figure \ref{fig:fft_sd} shows the steps executed for the FFT Poisson solver on the host and GPU. In case multiple host cores are sharing a GPU device, CUDA inter-process communications are used to share the device memory between the multiple MPI processes on the same node. For the FFT Poisson solver, one of the MPI processes acts as a main process and initialises memory allocation and kernel execution on the device. The other MPI processes meanwhile only send and receive data to and from the GPU. 

\subsubsection{Performance Results}
To test OPAL's performance, we use a similar problem setup as reported in \cite{Bi,NB}. The test
system consists of a host with two Intel Xeon e5-2609 v2 processors and a Nvidia Tesla K20 or Tesla K40. On the host, 8 CPU cores are available. The first simulations where run using only the CPUs available on the host. However in the second case, DKS is used to offload the FFT Poisson solver to the GPU.

\begin{table}[htb]
  \centering
  \small
  \caption{FFT Poisson Solver results}
  \begin{tabular}{c | c | c | C | C | C }
    FFT size & DKS & Total time (s) & OPAL speedup & FFTPoisson time (s) & FFTPoisson speedup\\
    \hline\hline
    
    \multirow{3}{*}{64x64x32} & no & 324.98 & & 22.53 & \\
    & K20 & \textbf{311.17} & 
    \textbf{\textcolor{red}{$\times$1.04}} & \textbf{7.42} & \textbf{\textcolor{red}{$\times$3}}\\
    & K40 & \textbf{293.7} & 
    \textbf{\textcolor{red}{$\times$1.10}} & \textbf{7.32} & \textbf{\textcolor{red}{$\times$3}}\\
    \hline

    \multirow{3}{*}{128x128x64} & no & 434.22 & & 206.73 & \\
    & K20 & \textbf{262.74} & 
    \textbf{\textcolor{red}{$\times$1.6}} & \textbf{32.15} & \textbf{\textcolor{red}{$\times$6.5}}\\
    & K40 & \textbf{245.08} & 
    \textbf{\textcolor{red}{$\times$1.8}} & \textbf{25.87} & \textbf{\textcolor{red}{$\times$8}}\\
    \hline
	
    \multirow{3}{*}{256x256x128} & no & 2308.05 & & 1879.84 & \\
    & K20 & \textbf{625.37} &
    \textbf{\textcolor{red}{$\times$3.6}} & \textbf{202.63} & \textbf{\textcolor{red}{$\times$9.3}}\\
    & K40 & \textbf{542.73} &
    \textbf{\textcolor{red}{$\times$4.2}} & \textbf{160.87} & \textbf{\textcolor{red}{$\times$11.7}}\\
    \hline

    \multirow{2}{*}{512x512x256} & no & 3760.46 & & 3327.14 & \\
    & K40 & \textbf{716.86} &
    \textbf{\textcolor{red}{$\times$5.2}} & \textbf{302.49} & \textbf{\textcolor{red}{$\times$11}}\\
        
  \end{tabular}

  \label{table:fftpoisson}

\end{table}

Table \ref{table:fftpoisson} shows the results of these test runs for multiple problem sizes. The results show that offloading FFT Poisson solver to GPU can provide a substantial speedup even when we have multiple CPU cores sharing one accelerator. The limiting factor for the performance of the FFT Poisson solver is the data transfer from the host side to the device. Since data needs to be moved to and from GPU at every time step, for the largest problem size reported in the benchmark tests, data transfer can take up to 55\% of the total simulation time. Another limiting factor is the performance of the FFT transform. FFT is a memory bound algorithm and is able to reach only a fraction (about 10\% was observed on our test system) of the devices peak performance. Since for the Poisson solver time to perform FFT takes up to 80\% of all time spent for calculations, the speed of the solver depends severely on the speed of the FFT.

\subsection{Particle Matter Interaction}
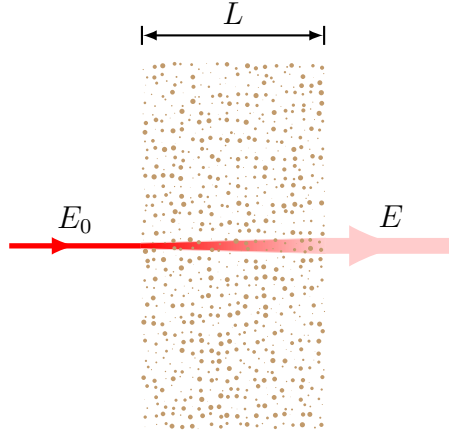
\begin{figure}
  \centering
  \begin{tikzpicture}[x          = \SquareUnit cm,
                    y          = \SquareUnit cm,
                    line width = 2pt
                   ]
\draw[red,
      decoration = {markings,
                    mark = at position 0.5 with {\arrow[]{latex}}},
      postaction = {decorate}] (-\BeforeLight,{\rows*\SquareUnit/2})--++
                               (\BeforeLight,0)node[midway,
                                                    above,
                                                    black]{$E_0$};


\draw[line width = 0pt, red!20!white, left color=red, right color=red!20!white] 
                   (0,{\rows*\SquareUnit/2+0.11}) -- 
                   (\cols*\SquareUnit,\rows*\SquareUnit/2+0.29) -- 
                   (\cols*\SquareUnit,\rows*\SquareUnit/2-0.29) -- 
                   (0,{\rows*\SquareUnit/2-0.11});

\draw[red!20!white, line width = 6pt,
      decoration = {markings,
                    mark = at position 0.6 with {\arrow[]{latex}}},
      postaction = {decorate}] ({\cols*\SquareUnit},{\rows*\SquareUnit/2})--++
                               (\BeforeLight,0)node[midway,
                                                    above,
                                                    black]{$E$};


\foreach \i in {1,...,\cols}{
	\foreach \j in {1,...,\rows}{
        \pgfmathsetmacro\radius{\RmaxParticle*\rnd}
        \pgfmathsetmacro\l{\SquareUnit-2*\radius}
        \pgfmathsetmacro\x{(\i-1)*\SquareUnit+\radius+\l*\rnd}
        \pgfmathsetmacro\y{(\j-1)*\SquareUnit+\radius+\l*\rnd}
        \fill[sand] (\x,\y)circle[radius=\radius];
        }
}

\draw[|<->|,
      >          = latex,
      line width = .8pt] ($(0,\rows*\SquareUnit)+(0,1)$)--++
                         (\cols*\SquareUnit,0)node[midway,
                                                   above]{$L$};

\end{tikzpicture}
\caption{Particle matter interaction. With the final energy $E<E_0$ and larger momenta spread due to  Coulomb scattering and the large angle Rutherford scattering.} 
\label{fig:degrader}
\end{figure}

One of the features in OPAL is the ability to perform Monte Carlo simulations of the particle beam interaction with matter. A fast charged particle moving through the material undergoes collisions with the atomic electrons and loses energy. In addition, particles are also deflected from their original trajectory due to the Coulomb scattering with nuclei, as shown in figure \ref{fig:degrader}. The energy loss in OPAL is calculated using Bethe-Bloch formula and the change of particle trajectory is simulated using Multiple Coulomb Scattering and Single Rutherford Scattering \cite{ICRU,PDG,William}. At every time step when the particle beam is inside a material, the following steps are executed:
\begin{itemize}
  \item calculate the energy loss of the beam,
  \item delete the particle if the particle's kinetic energy is smaller than a given threshold, 
  \item apply Coulomb \& Rutherford scattering to the beam. 
\end{itemize}

\subsubsection{The Energy Loss}

The energy loss is calculated using the following Bethe-Bloch equation:

\begin{equation}
\label{eq:dEdx}
-dE/dx=\frac{K z^2 Z}{A \beta^2}\left[\frac{1}{2} \ln{\frac{2 m_e c^2\beta^2 \gamma^2 Tmax}{I^2}}-\beta^2 \right],
\end{equation}
where $Z$ is the atomic number of absorber, $A$ is the atomic mass of absorber, $m_e$ is the electron mass, and $z$ is the charge number of the incident particle. $K$ is defined as $4\pi N_Ar_e^2m_ec^2$, where $r_e$ is the classical electron radius, $N_A$ is the Avogadro's number, and $I$ is the mean excitation energy. $\beta$ and $\gamma$ are the kinematic variables. Lastly $T_{max}$ is the maximum kinetic energy that can be imparted to a free electron in a single collision. It is defined as,
\begin{equation}
T_{max}=\frac{2m_ec^2\beta^2\gamma^2}{1+2\gamma m_e/M+(m_e/M)^2},
\end{equation}
where $M$ is the incident particle mass. 

For relatively thick absorbers, the number of collisions is large, and therefore the energy loss distribution is Gaussian in form.
For non-relativistic heavy particles, the spread, $\sigma_0$, of the Gaussian distribution is calculated by:
\begin{equation}
\sigma_0^2=4\pi N_Ar_e^2m_e^2c^4\rho\frac{Z}{A}\Delta s,
\end{equation}
where $\rho$ is the density and $s$ is the thickness of the material.

\subsubsection{Coulomb Scattering}
The Coulomb scattering is treated as two independent events: the multiple Coulomb scattering and the large angle Rutherford scattering.

Using the distribution given in \cite{Jackson}, the multiple- and single-scattering distributions can be written as:

\begin{eqnarray}
\label{eq:PM}
P_M(\alpha) d \alpha &=& \frac{1}{\sqrt{\pi}}e^{-\alpha^2}d\alpha, \\ 
\label{eq:Ps}
P_S(\alpha) d \alpha & =& \frac{1}{8 \ln(204 Z^{-1/3})} \frac{d \alpha}{\alpha^3},
\end{eqnarray}

where $\alpha=\frac{\theta}{\langle\Theta^2\rangle^{1/2}}=\frac{\theta}{\sqrt 2 \theta_0}$.
The transition point between multi and single scattering occurs at the angle $\theta=2.5 \sqrt 2 \theta_0\approx3.5 \theta_0$, where value of $\theta_0$ is the scattering angle from Moliere's theory and is defined as,
\begin{equation}
\label{eq:theta0}
\theta_0=\frac{13.6MeV}{\beta c p} z \sqrt{\Delta s/X_0} [1+0.038 \ln(\Delta s/X_0)],
\end{equation}
where $p$ is the momentum, $\Delta s$ is the step size, and $X_0$ is the radiation length. 

To perform a Monte Carlo simulation for the multiple Coulomb scattering two independent Gaussian random variables ($z_1$ and $z_2$) are created with mean zero and variance one. The new position and momentum can then be calculated by:
\begin{eqnarray}
\label{eq:Multiplex}
x &=& x+\Delta s p_x+z_1 \Delta s \theta_0/\sqrt{12}+z_2 \Delta s \theta_0/2, \\
\label{eq:Multiplepx}
p_x &=& p_x+z_2 \theta_0.
\end{eqnarray}
The values for the $y-p_y$ plane are calculated with the very same Monte-Carlo algorithm. 


\subsubsection{Large Angle Rutherford Scattering}

Only a small percentage of particles undergo large angle Rutherford scattering. This percentage is given by:

\begin{equation}
\label{eq:rutherfordscat}
\chi_{single}<\frac{\int_{2.5}^\infty P_S(\alpha)d\alpha}{\int_0^{2.5} P_M(\alpha)d\alpha+\int_{2.5}^\infty P_S(\alpha)d\alpha} = 0.0047.
\end{equation}

The process to define if a particle undergoes a Rutherford scatter is as follows:
\begin{itemize}
\item A random number $\xi_1$ between 0 and 1 is generated. If and only if this random number is smaller than $\chi_{single}$ the particle undergoes single Rutherford scattering. The value of $\chi_{single}$ does not change significantly for different materials, hence a fixed value of $\chi_{single}=0.0047$ is used, in order to avoid unnecessary computation.

\item A second random variable $\xi_2$ between 0 and 1 is generated to calculate the angle, the particle rotates about.


\item The third and last random number $\xi_3$ determines the direction of the rotation:
\end{itemize}

\begin{equation}
  \theta_{Ru} = \begin{cases} +2.5 \sqrt{\frac{1}{\xi_2}} \sqrt{2}\theta_0 & \text{if } \quad \xi_3 < 0.5 \\
-2.5 \sqrt{\frac{1}{\xi_2}} \sqrt{2}\theta_0 & \text{if } \quad \xi_3 > 0.5. \end{cases}
\end{equation}


\subsubsection{The DKS Implementation of the Particle Matter Interaction Model}
\begin{figure}[htb]
  \centering
\begin{tikzpicture}[scale=0.6, every node/.style={transform shape, font=\normalsize,
  minimum height=0.5cm,minimum width=0.5cm},]

\def\CPU{Host}
\def\GPU{Stream 1}
\def\GGPU{Stream 2}

\node [matrix, very thin,column sep=0.4cm,row sep=0.3cm] (matrix) at (0,0) {
  & \node(0,0) (\CPU) {}; & \node(0,0) (\CPU 0) {}; & \node(0,0) (\CPU 1) {}; & 
  \node(0,0) (\CPU 2) {}; & \node(0,0) (\CPU 3) {}; & \node(0,0) (\CPU 4) {}; & 
  \node(0,0) (\CPU 5) {}; & \node(0,0) (\CPU 6) {}; & \node(0,0) (\CPU 7) {}; & 
  \node(0,0) (\CPU 8) {}; & \node(0,0) (\CPU 9) {}; & \node(0,0) (\CPU 10) {};\\
  & & & & & & & & & & &\\
  & \node(0,0) (\GPU) {}; & \node(0,0) (\GPU 0) {}; & \node(0,0) (\GPU 1) {}; & 
  \node(0,0) (\GPU 2) {}; & \node(0,0) (\GPU 3) {}; & \node(0,0) (\GPU 4) {}; & 
  \node(0,0) (\GPU 5) {}; & \node(0,0) (\GPU 6) {}; & \node(0,0) (\GPU 7) {}; & 
  \node(0,0) (\GPU 8) {}; & \node(0,0) (\GPU 9) {}; & \node(0,0) (\GPU 10) {};\\
  & & & & & & & & & & &\\
  & \node(0,0) (\GGPU) {}; & \node(0,0) (\GGPU 0) {}; & \node(0,0) (\GGPU 1) {}; & 
  \node(0,0) (\GGPU 2) {}; & \node(0,0) (\GGPU 3) {}; & \node(0,0) (\GGPU 4) {}; & 
  \node(0,0) (\GGPU 5) {}; & \node(0,0) (\GGPU 6) {}; & \node(0,0) (\GGPU 7) {}; & 
  \node(0,0) (\GGPU 8) {}; & \node(0,0) (\GGPU 9) {}; & \node(0,0) (\GGPU 10) {};\\
};

\fill 
(\CPU) node[draw,fill=white] {\textbf{\CPU}}
(\GPU) node[draw,fill=white] {\textbf{\GPU}}
(\GGPU) node[draw,fill=white] {\textbf{\GGPU}};

\filldraw[fill=blue!20]
(\CPU 0.north west) rectangle (\CPU 4.south east) node[pos=.5] {Initialize, data transfer, launch kernels}
(\CPU 7.north west) rectangle (\CPU 8.south east) node[pos=.5] {Receive data}
(\GPU 0.north west) rectangle (\GPU 0.south east) node[pos=.5] {I}

(\GGPU 1.north west) rectangle (\GGPU 2.south east) node[pos=.5] {Receive data}
(\GGPU 4.north west) rectangle (\GGPU 5.south east) node[pos=.5] {Integrate}
(\GGPU 7.north west) rectangle (\GGPU 8.south east) node[pos=.5] {Send data}

(\GPU 2.north west) rectangle (\GPU 3.south east) node[pos=.5] {Receive data}
(\GPU 5.north west) rectangle (\GPU 6.south east) node[pos=.5] {Integrate}
(\GPU 8.north west) rectangle (\GPU 9.south east) node[pos=.5] {Send data};

\filldraw
(\CPU 9) node[draw,circle,fill=red!20] {$R, X$}
(\GGPU 3) node[draw,circle,fill=green!20] {$X, dt$}
(\GPU 4) node[draw,circle,fill=green!20] {$R, P$}
(\GGPU 6) node[draw,circle,fill=green!20] {$X$}
(\GPU 7) node[draw,circle,fill=green!20] {$R$};

\draw [dashed] (\CPU 4) -- (\CPU 7);
\draw [dashed] (\CPU 9) -- (\CPU 10);
\draw [-latex] (\CPU 8) -- (\CPU 9);

\draw [-latex] (\GPU 0) -- (\GPU 2);
\draw [-latex] (\GPU 3) -- (\GPU 4);
\draw [-latex] (\GPU 4) -- (\GPU 5);
\draw [-latex] (\GPU 6) -- (\GPU 7);
\draw [-latex] (\GPU 7) -- (\GPU 8);
\draw [dashed] (\GPU 9) -- (\GPU 10);

\draw [dashed] (\GGPU 0) -- (\GGPU 1);
\draw [-latex] (\GGPU 2) -- (\GGPU 3);
\draw [-latex] (\GGPU 3) -- (\GGPU 4);
\draw [-latex] (\GGPU 5) -- (\GGPU 6);
\draw [-latex] (\GGPU 6) -- (\GGPU 7);
\draw [dashed] (\GGPU 8) -- (\GGPU 10);

\draw [-latex] (\CPU 0) -- (\GPU 0);
\draw [-latex] (\CPU 1) -- (\GGPU 1);
\draw [-latex] (\CPU 2) -- (\GPU 2);
\draw [shorten <=0.25cm,shorten >=0.25cm,-latex] ($(\GGPU 7)!0.4!(\GGPU 8)$) -- ($(\CPU 7)!0.4!(\CPU 8)$);
\draw [-latex] (\GPU 7) -- (\CPU 7);
\draw [-latex] (\GPU 4) -- (\GGPU 4);

\end{tikzpicture}
  \caption{Integration sequence diagrams}
  \label{fig:integration_sd}
\end{figure}
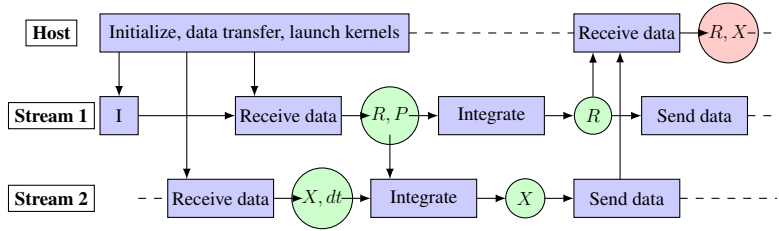

For particle matter interactions, DKS has CUDA and OpenMP implementations of all the algorithm steps described above. This allows the computation of the energy loss, the Coulomb scattering, and the Rutherford scattering to be offloaded to the GPU or Intel MIC. On top of particle matter interaction, DKS is also able to offload to the accelerators the transport of particles before and after the material using a time integration scheme. The sequence diagram for the integration is shown in Figure \ref{fig:integration_sd}.

To increase the performance, the data transfer is minimised as much as possible and particles that are drifting before or after the degrader are kept on the device. They are updated only when there are a some particles returning from the material or there has been an MPI update to balance the workload between MPI processes. Pinned host memory and streams are used with the GPU version to increase the date transfer speed, and overlap the data transfer and kernel execution for the particle drift.

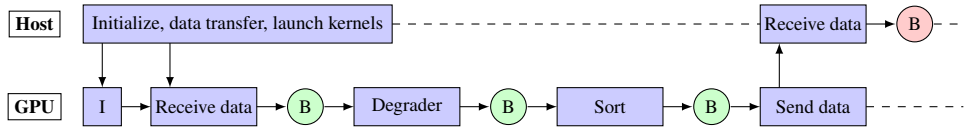
\begin{figure}[htb]
  \centering
\begin{tikzpicture}[scale=0.6, every node/.style={transform shape, font=\normalsize,
  minimum height=0.5cm,minimum width=0.5cm},]

\def\CPU{Host}
\def\GPU{GPU}

\node [matrix, very thin,column sep=0.4cm,row sep=0.3cm] (matrix) at (0,0) {
  & \node(0,0) (\CPU) {}; & \node(0,0) (\CPU 0) {}; & \node(0,0) (\CPU 1) {}; & 
  \node(0,0) (\CPU 2) {}; & \node(0,0) (\CPU 3) {}; & \node(0,0) (\CPU 4) {}; & 
  \node(0,0) (\CPU 5) {}; & \node(0,0) (\CPU 6) {}; & \node(0,0) (\CPU 7) {}; & 
  \node(0,0) (\CPU 8) {}; & \node(0,0) (\CPU 9) {}; & \node(0,0) (\CPU 10) {}; & 
  \node(0,0) (\CPU 11) {}; & \node(0,0) (\CPU 12) {}; & \node(0,0) (\CPU 13) {}; &
  \node(0,0) (\CPU 14) {}; & \node(0,0) (\CPU 15) {}; & \node(0,0) (\CPU 16) {};\\
  & & & & & & & & & & & & & & & & & &\\ 
  & \node(0,0) (\GPU) {}; & \node(0,0) (\GPU 0) {}; & \node(0,0) (\GPU 1) {}; & 
  \node(0,0) (\GPU 2) {}; & \node(0,0) (\GPU 3) {}; & \node(0,0) (\GPU 4) {}; & 
  \node(0,0) (\GPU 5) {}; & \node(0,0) (\GPU 6) {}; & \node(0,0) (\GPU 7) {}; & 
  \node(0,0) (\GPU 8) {}; & \node(0,0) (\GPU 9) {}; & \node(0,0) (\GPU 10) {}; & 
  \node(0,0) (\GPU 11) {}; & \node(0,0) (\GPU 12) {}; & \node(0,0) (\GPU 13) {}; &
  \node(0,0) (\GPU 14) {}; & \node(0,0) (\GPU 15) {}; & \node(0,0) (\GPU 16) {};\\
};

\fill 
(\CPU) node[draw,fill=white] {\textbf{\CPU}}
(\GPU) node[draw,fill=white] {\textbf{\GPU}};

\filldraw[fill=blue!20]
(\CPU 0.north west) rectangle (\CPU 4.south east) node[pos=.5] {Initialize, data transfer, launch kernels}
(\GPU 0.north west) rectangle (\GPU 0.south east) node[pos=.5] {I}
(\GPU 1.north west) rectangle (\GPU 2.south east) node[pos=.5] {Receive data}
(\GPU 4.north west) rectangle (\GPU 5.south east) node[pos=.5] {Degrader}
(\GPU 7.north west) rectangle (\GPU 8.south east) node[pos=.5] {Sort}
(\GPU 10.north west) rectangle (\GPU 11.south east) node[pos=.5] {Send data}
(\CPU 10.north west) rectangle (\CPU 11.south east) node[pos=.5] {Receive data};

\draw
(\CPU 12) node[draw,circle,fill=red!20] {B}
(\GPU 3) node[draw,circle,fill=green!20] {B}
(\GPU 6) node[draw,circle,fill=green!20] {B}
(\GPU 9) node[draw,circle,fill=green!20] {B};

\draw [dashed] (\CPU 4) -- (\CPU 10);
\draw [-latex] (\CPU 11) -- (\CPU 12);
\draw [dashed] (\CPU 12) -- (\CPU 13);

\draw [-latex] (\GPU 0) -- (\GPU 1);
\draw [-latex] (\GPU 2) -- (\GPU 3);
\draw [-latex] (\GPU 3) -- (\GPU 4);
\draw [-latex] (\GPU 5) -- (\GPU 6);
\draw [-latex] (\GPU 6) -- (\GPU 7);
\draw [-latex] (\GPU 8) -- (\GPU 9);
\draw [-latex] (\GPU 9) -- (\GPU 10);
\draw [dashed] (\GPU 11) -- (\GPU 13);

\draw [-latex] (\CPU 0) -- (\GPU 0);
\draw [-latex] (\CPU 1) -- (\GPU 1);
\draw [-latex] (\GPU 10) -- (\CPU 10); 

\end{tikzpicture}
  \caption{Degrader sequence diagrams, where B denotes particle bunch}
  \label{fig:degrader_sd}
\end{figure}

Particles that are in the material are also kept in the device memory. NVIDIAs cuRAND and Intels MKL VSL libraries are used to generate random numbers to determine the necessary distributions for energy loss and scattering. NVIDIA's Thrust library is used to sort and count the particles on the GPU in order to manage the particles that need to come out of the material, but also to exclude the dead particles from Monte-Carlo simulations. Because of the high complexity of the algorithm, the CUDA version uses shared device memory for variable storage to reduce the register pressure of the kernels in order to achieve higher GPU occupancy. Structure of arrays data layout is used to store all the particles in order to allow Intel compiler to better vectorise the code for the Xeon Phi coprocessor. The sequence diagram of the degrader simulations on the CPU and the device is shown in Figure \ref{fig:degrader_sd}.

\subsubsection{Performance Results}
To test the OPAL and DKS Monte Carlo simulations, an example was run where particles are shot through  a $L=1$ cm thick graphite slab. This mimics a degrader device used in proton therapy. Timings where obtained for the integration of the equation of motion, before and after the material, as well as when the particles are moving through the material. The system setup is the same as was used for the FFT Based Particle-Mesh (PM) Solver benchmark, but, because of limitations of OPAL, simulations where run using only one CPU core.

Table \ref{table:degrader} shows the benchmark results for the degrader and the integration using various number of particles. The speedup of the particle transport through the material is $\times 140$ to $\times 170$ compared to the one core of the host processor, while the integration is able to achieve a speedup of around $\times 20$ to $\times 30$ on the GPU. The Intel MIC on the other hand shows a speedup of  $\times 40$ for the degrader and $\times 7$ for the integration compared to the host.


\begin{table}[htb]
  \centering
  \small
  \caption{OPAL degrader results}
  \begin{tabular}{c | c | c | C | C | C | C}
    Particles & DKS & Degrader time (s) & Degrader speedup & Integration time (s) & Integration speedup\\
    \hline\hline
    \multirow{4}{*}{$10^5$} & no & 20.30 & & 3.46 & \\
    & MIC & \textbf{2.29} & \textbf{\textcolor{red}{$\times$8}} & \textbf{0.89} & \textbf{\textcolor{red}{$\times$4}}\\
    & K20 & \textbf{0.28} & \textbf{\textcolor{red}{$\times$72}} & \textbf{0.15} & \textbf{\textcolor{red}{$\times$23}}\\
    & K40 & \textbf{0.19} & \textbf{\textcolor{red}{$\times$107}} & \textbf{0.14} & \textbf{\textcolor{red}{$\times$24}}\\
    \hline
    \multirow{4}{*}{$10^6$} & no & 206.77 & & 34.93 & \\
    & MIC & \textbf{5.38} & \textbf{\textcolor{red}{$\times$38}} & \textbf{4.62} & \textbf{\textcolor{red}{$\times$7.5}}\\
    & K20 & \textbf{1.41} & \textbf{\textcolor{red}{$\times$146}} & \textbf{1.83} & \textbf{\textcolor{red}{$\times$19}}\\
    & K40 & \textbf{1.18} & \textbf{\textcolor{red}{$\times$175}} & \textbf{1.21} & \textbf{\textcolor{red}{$\times$29}}\\
    \hline
    \multirow{3}{*}{$10^7$} & no & 2048.25 & & 351.64 & \\
    & K20 & \textbf{14.4} & \textbf{\textcolor{red}{$\times$142}} & \textbf{17.21} & \textbf{\textcolor{red}{$\times$20}}\\
    & K40 & \textbf{12.79} & \textbf{\textcolor{red}{$\times$160}} & \textbf{11.43} & \textbf{\textcolor{red}{$\times$30}}\\
  \end{tabular}    
  \label{table:degrader}
\end{table}

The limiting factor for GPU/MIC performance for the integration is the data movement. This operation requires data to be sent to the device and received from the device, at every time step. In the GPU case kernel execution can be completely overlapped with the data movement, and thus only limited by the memory bandwidth of our device.

The limiting factor for particle movement through the material is global memory access times. Each execution of the kernel requires a load of position and momentum vectors as well as the state of the random number generator for the thread. When the kernel finishes position, momentum, random number state and possibly particle state needs to be written back to global memory. If there are any dead particles, or particles that are coming out of the material, these particles need to be removed from the bunch on the accelerator. This requires a sorting of the particles which also requires a lot of memory movement on the device and limits the performance.

\section{Parameter Fitting with \textsc{Minuit2}}
\label{sec:paramfit}

\textsc{Minuit2} is a \texttt{C++} library allowing a multi-parameter minimisation of a user-defined function \cite{moneta2005developments}. It is a re-implementation of the FORTRAN library \textsc{Minuit} \cite{minuit}, a very popular minimisation package used by high energy physicists. In addition to minimisation algorithms, it contains methods for analysing the solutions and can estimate the parameter error correlation matrix. These combined capabilities are very difficult to find in other existing minimisers. Its drawback for inexperienced users is that the user-defined function needs to be implemented, compiled, and linked. This is a common practice in high energy physics, but is less common in the solid state physics community. Therefore, for the $\mu$SR community, the \textsc{Musrfit} framework \cite{suter2012musrfit} has been developed. This framework eases the analysis of muon spin rotation, relaxation, and resonance ($\mu$SR) experiments by allowing user to define all the relevant input parameters and functions for \textsc{Minuit2} in a scripting manner. 
We will describe the problem for the specific needs of $\mu$SR, however the problem and the described solution is much more generic. 

\subsection{Problem Description}

In a time differential $\mu$SR experiment \cite{youanc2011muon}, $\sim100$\% polarised positive muons ($\mu^+$) are implanted in a solid sample and rapidly thermalise ($\sim 10$ ps) without noticeable polarisation loss. The spin evolution of the muon ensemble after implantation is then measured as a function of time. The evolution can be monitored by using the fact that the parity violating muon decay is highly anisotropic. During the decay an easily detectable positron is emitted preferentially in the direction of the $\mu^+$ spin. It takes the form

\begin{equation}\label{eq:positron_histo}
 N^j(t) = N_0^j e^{-t/\tau_\mu} [ 1 + A^j({\vec p}^j, t) ] + N_{\rm bkg}^j,
\end{equation}

\noindent where the time is measured in discrete steps $t = n \cdot \Delta t$ [$n \in \mathbb{N}_0$, $\Delta t$ the time resolution]. $j$ indexes the positron detectors and the function $A^j({\vec p}, t)$ describes the ``physics'' of the system under consideration. For details about the function $A^j({\vec p}, t)$, the reader is referred to Ref.\ \cite{youanc2011muon}. The muon lifetime is given by $\tau_\mu$ and  $N_0$ gives the scale of the positron count. Lastly the constant $N_{\rm bkg}^j$ originates from uncorrelated background events. For a given positron histogram, $j$, the optimal parameter set 

\begin{equation}
 {\vec P}^j = \left\{ N_0^j, N_{\rm bkg}^j, {\vec p}^j \right\} 
\end{equation}

\noindent needs to be determined. Depending on the level of statistics of the positron histograms, ${\vec P}$ is determined with a $\chi^2$ minimisation:

\begin{equation}\label{eq:chisq}
 \chi^2({\vec P}) = \sum_j \sum_n \frac{[d_n^j - N^j(n\cdot\Delta t, {\vec P})]^2}{(d_{n, \rm err}^j)^2},
\end{equation}

\noindent where $d_n^j$ are the measured data points of $j^{\rm th}$ positron detector. The theory describing the data is given by Eq.\ (\ref{eq:positron_histo}), and $d_{n, \rm err}^j$ is the estimated error of $d_n^j$ ($d_{n, \rm err}^j = \sqrt{d_n^j}$ for the Poisson distributed positron histogram). 

For data sets with rather limited statistics, Eq.\ (\ref{eq:chisq}) is not leading to satisfactory results. In this case the log-likelihood function

\begin{equation}\label{eq:log-likelihood}
  {\cal L} = 2 \cdot \sum_j \sum_n 
            \begin{cases} 
               \left[ N^j(n\cdot\Delta t, {\vec P}) - d_n^j \right] + d_n^j \log\left[\frac{\displaystyle d_n^j}{\displaystyle N^j(n\cdot\Delta t, {\vec P})}\right], & d_n^j > 0 \\
               \left[ N^j(n\cdot\Delta t, {\vec P}) - d_n^j \right] & d_n^j \leq 0
            \end{cases}
\end{equation}

\noindent should be maximised and lead to much better estimates of ${\vec P}$.

In recent years, changes in the detector technology allow higher time resolution (smaller detectable $\Delta t$) at the expense of higher detector fragmentation (more positron counters). This is leading to much larger data sets, and as a consequence the minimisation/maximisation times are exploding. 
This is especially true if a parameter error estimate is needed that goes beyond the simple Hessian approach.

The \textsc{Minuit2} library is almost perfectly suited to tackle the problem. The user needs to implement Eqs.\ (\ref{eq:chisq}) and/or (\ref{eq:log-likelihood}) on the accelerator(s), whereas the minimisation process is executed on the host. The main, and most time consuming, part, in the calculation is given by Eqs.\ (\ref{eq:chisq}) and (\ref{eq:log-likelihood}). The only data transfer needed between the host and the accelerator is given by the small parameter set ${\vec P}$, which should not lead to a bottleneck in the overall computation time.

\subsubsection{The DKS Implementation}
For parameter fitting with \textsc{Minuit2}, DKS is used to offload the $\chi^2$, max-log-likelihood, and the user defined function calculations to the GPU. The calculated value is passed to \textsc{Minuit2} which varies the parameter set and returns the new parameters. Data transfer from the host to the device is minimal. Measurement data are transferred to the GPU only at the beginning of the calculations. Only the small parameter array is transferred to the device before every calculation and only the $\chi^2$ or max-log-likelihood value is returned by the device. CUDA and OpenCL are implemented to support various hardware accelerators.

\subsubsection{Performance Results}
Parameter fitting tests with \textsc{Minuit2} and DKS were performed on the same machine as OPAL tests. The host code used 8 CPU cores and parallelisation was done using OpenMP. The device code was run on a Nvidia Tesla K20 and Tesla K40 using CUDA. Results for offloading $\chi^2$ and max-log-likelihood functions are shown in the Table \ref{table:chisquare}. Since there is almost no data transfer involved in the program and the algorithm is easy to parallelise it is a good candidate for GPU acceleration. The time to solution on the GPU is around 300 to 400 times faster for $\chi^2$ and around 150 to 200 times faster for max-log-likelihood functions than currently used OpenMP implementation. From the results presented in Tab. \ref{tab:minuit2-results}, $A^{j}({\vec p},t)$ was chosen as

\begin{equation}\label{eq:asymmetry}
 A^{j}({\vec p},t) = A_0^j \exp\left[-\frac{1}{2} (\sigma t)^2\right] \, \cos(\gamma_\mu B t + \phi^j),
\end{equation}

\noindent with $j=1\ldots 16$, $A_0^j$ are the asymmetries of each positron detector, $\sigma$ the depolarisation rate of the muon spin ensemble, $\gamma_\mu$ the gyromagnetic ratio of the muon, $B$ the magnetic induction at the muon stopping site, $t$ the time [see Eq.(\ref{eq:positron_histo})], and $\phi^j$ the phase of the initial muon spin in respect to the positron detector. Eq.(\ref{eq:asymmetry}) is a typical muon polarisation function to determine the magnetic shift of a para-/diamagnetic material (see Ref. \cite{youanc2011muon}). For the given number of positron detectors, 66 fitting parameters needed to be determined [see Eqs.(\ref{eq:positron_histo}), (\ref{eq:asymmetry})].

\begin{table}[h]
  \caption{Minuit2 parameter fitting with $\chi^2$ and max-log-likelihood (MLE) function running on the GPU.
           The given time is for the execution of the \texttt{migrad} command of Minuit2 \cite{moneta2005developments}.}\label{tab:minuit2-results}
  \centering
  \small{
    \begin{tabular}{c | c | c | c | c | c}
      Data Set Size & DKS & $\chi^2$ (s) & Speedup & MLE (s) & Speedup\\
      \hline\hline
      \multirow{3}{*}{$\sim$1,300,000} & no & 157.077 & & 446.444 &\\
      & K20 & \textbf{0.55012} & \textbf{\textcolor{red}{$\times$285}} & \textbf{2.75018} & \textbf{\textcolor{red}{$\times$162}}\\
      & K40 & \textbf{0.432707} & \textbf{\textcolor{red}{$\times$363}} & \textbf{2.31758} & \textbf{\textcolor{red}{$\times$193}}\\
      \hline
      \multirow{3}{*}{$\sim$1,700,000} & no & 264.279 & & 664.893 &\\
      & K20 & \textbf{0.863937} & \textbf{\textcolor{red}{$\times$306}} & \textbf{4.16143} & \textbf{\textcolor{red}{$\times$160}} \\
      & K40 & \textbf{0.675295} & \textbf{\textcolor{red}{$\times$391}} & \textbf{3.48775} & \textbf{\textcolor{red}{$\times$190}} \\
     \hline
      \multirow{3}{*}{$\sim$2,200,000} & no & 392.727 & & 741.114 &\\
      & K20 & \textbf{1.32307} & \textbf{\textcolor{red}{$\times$296}} & \textbf{5.2768} & \textbf{\textcolor{red}{$\times$140}} \\
      & K40 & \textbf{1.02269} & \textbf{\textcolor{red}{$\times$384}} & \textbf{4.41817} & \textbf{\textcolor{red}{$\times$168}} \\

      \hline
      \multirow{3}{*}{$\sim$3,300,000} & no & 859.339 & & 1101.62 &\\
      & K20 & \textbf{2.52918} & \textbf{\textcolor{red}{$\times$339}} & \textbf{7.60934} & \textbf{\textcolor{red}{$\times$144}} \\
      & K40 & \textbf{1.92734} & \textbf{\textcolor{red}{$\times$446}} & \textbf{6.30366} & \textbf{\textcolor{red}{$\times$175}}
      
    \end{tabular}
  }
  \label{table:chisquare}
\end{table}

\section{Conclusion and Outlook}
In this paper we presented the first version of the Dynamic Kernel Scheduler which provides a software layer between the host application and the hardware accelerators. This creates a fine tuned code for different hardware accelerators using different frameworks and allows to easily integrate it into existing host applications. DKS was integrated into OPAL to offload FFT based Poisson solver and Monte-Carlo simulations for particle-matter interactions to a GPU and a Intel MIC using either CUDA or OpenMP. DKS was also used together with Minuit2 for parameter fitting, where $\chi^2$ and max-log-likelihood function calculations were offloaded to a GPU using CUDA or OpenCL. The result of this work shows that DKS can be used to substantially speed up existing host applications with minimal additions and changes to host code. Separating the device specific code in a different layer allows managing and fine tuning of the code more easily and keeps the host application significantly more portable since all the device and framework specific details are handled by DKS.

\label{sec:concloutl}

\section{References}

\bibliographystyle{elsarticle-num}  
 \bibliography{dksbib}{}

\begin{thebibliography}{10}
\expandafter\ifx\csname url\endcsname\relax
  \def\url#1{\texttt{#1}}\fi
\expandafter\ifx\csname urlprefix\endcsname\relax\def\urlprefix{URL }\fi
\expandafter\ifx\csname href\endcsname\relax
  \def\href#1#2{#2} \def\path#1{#1}\fi

\bibitem{top500}
Top500 supercomputers, http://www.top500.org/lists/2015/06/.

\bibitem{cuda}
{NVIDIA} {CUDA} {Z}one, https://developer.nvidia.com/cuda-zone.

\bibitem{opencl}
Khronos {G}roup, {O}pen{CL}, https://www.khronos.org/opencl/.

\bibitem{mic}
Intel {X}eon {P}hi coprocessor, https://software.intel.com/en-us/mic-developer.

\bibitem{openacc}
Open{ACC}, http://www.openacc.org/.

\bibitem{openmp}
Open{MP}, http://openmp.org/wp/openmp-specifications/.

\bibitem{Decyk2011}
V.~K. Decyk, T.~V. Singh, {Adaptable Particle-in-Cell algorithms for graphical
  processing units}, Computer Physics Communications 182~(3) (2011) 641--648.

\bibitem{Rossinelli2010}
D.~Rossinelli, M.~Bergdorf, G.-H. Cottet, P.~Koumoutsakos, {GPU accelerated
  simulations of bluff body flows using vortex particle methods}, Journal of
  Computational Physics 229~(9) (2010) 3316--3333.

\bibitem{Dugan2013}
N.~Dugan, L.~Genovese, S.~Goedecker, {A customized 3D GPU Poisson solver for
  free boundary conditions}, Computer Physics Communications 184~(8) (2013)
  1815--1820.

\bibitem{Wu2013}
J.~Wu, J.~JaJa, {High performance FFT based poisson solver on a CPU-GPU
  heterogeneous platform}, Proceedings - IEEE 27th International Parallel and
  Distributed Processing Symposium, IPDPS 2013 (2013) 115--125.

\bibitem{Wu2014}
J.~Wu, J.~JaJa, {Optimized FFT computations on heterogeneous platforms with
  application to the Poisson equation}, Journal of Parallel and Distributed
  Computing 74~(8) (2014) 2745--2756.

\bibitem{Osiecki2013}
T.~H. Osiecki, M.-y. Tsai, A.~E. Gattiker, D.~A. Jamsek, S.~R. Nassif, W.~E.
  Speight, C.~C. Sze, {Hardware Acceleration of an Efficient and Accurate
  Proton Therapy Monte Carlo}, Procedia Computer Science 18 (2013) 2241--2250.

\bibitem{Jia2012}
X.~Jia, T.~Pawlicki, K.~T. Murphy, A.~J. Mundt, {Proton therapy dose
  calculations on GPU: advances and challenges}, Translational Cancer Research
  1~(3) (2012) 207--216.

\bibitem{Jia2011}
X.~Jia, X.~Gu, Y.~J. Graves, M.~Folkerts, S.~B. Jiang, {GPU-based fast Monte
  Carlo simulation for radiotherapy dose calculation} (2011) 18.

\bibitem{Liu2015}
T.~Liu, X.~Xu, C.~Carothers, {Comparison of two accelerators for Monte Carlo
  radiation transport calculations, Nvidia Tesla M2090 GPU and Intel Xeon Phi
  5110p coprocessor: A case study for X-ray CT imaging dose calculation},
  Annals of Nuclear Energy 82 (2015) 230--239.

\bibitem{Xu2015}
X.~G. Xu, T.~Liu, L.~Su, X.~Du, M.~Riblett, W.~Ji, D.~Gu, C.~D. Carothers,
  M.~S. Shephard, F.~B. Brown, M.~K. Kalra, B.~Liu, {ARCHER, a new Monte Carlo
  software tool for emerging heterogeneous computing environments}, Annals of
  Nuclear Energy 82 (2015) 2--9.

\bibitem{Tickner2010}
J.~Tickner, {Monte Carlo simulation of X-ray and gamma-ray photon transport on
  a graphics-processing unit}, Computer Physics Communications 181~(11) (2010)
  1821--1832.

\bibitem{nvidialibraries}
{GPU} {A}ccelerated {L}ibraries,
  https://developer.nvidia.com/gpu-accelerated-libraries.

\bibitem{miclibraries}
Intel {M}ath {K}ernel {L}ibrary ({I}ntel {MKL}),
  https://software.intel.com/en-us/articles/intel-mkl-on-the-intel-xeon-phi-coprocessors.

\bibitem{Bourgoin2014}
M.~Bourgoin, E.~Chailloux, J.~L. Lamotte, {Efficient abstractions for GPGPU
  programming}, International Journal of Parallel Programming 42~(4) (2014)
  583--600.

\bibitem{Svensson2010}
J.~Svensson, K.~Claessen, M.~Sheeran, {GPGPU kernel implementation and
  refinement using Obsidian}, Procedia Computer Science 1~(1) (2010)
  2065--2074.

\bibitem{Vinas2015}
M.~Vi\~{n}as, B.~B. Fraguela, Z.~Bozkus, D.~Andrade, {Improving OpenCL
  Programmability with the Heterogeneous Programming Library}, Procedia
  Computer Science 51 (2015) 110--119.

\bibitem{hockney1970methods}
R.~Hockney, Methods in computational physics, Alder, B (1970) 136--211.

\bibitem{eastwoodandbrownrigg}
J.~W. Eastwood, D.~R.~K. Brownrigg, J. Comp. Phys, 32, 24-38.

\bibitem{Hockney}
R.~Hockney, J.~Eastwood, Computer Simulation using Particles, Adam Hilger,
  1988.

\bibitem{Bi}
Y.~Bi, A.~Adelmann, R.~D\"{o}lling, M.~Humbel, W.~Joho, M.~Seidel, T.~Zhang,
  Towards quantitative simulations of high power proton cyclotrons, Physical
  Review Special Topics - Accelerators and Beams 14~(5).

\bibitem{NB}
J.~Yang, A.~Adelmann, M.~Humbel, M.~Seidel, T.~Zhang, Beam dynamics in high
  intensity cyclotrons including neighboring bunch effects: Model,
  implementation, and application, Physical Review Special Topics -
  Accelerators and Beams 13~(6).

\bibitem{ICRU}
Stopping powers and ranges for protons and alpha particles, ICRU Report 49.

\bibitem{PDG}
{K.A. Olive et al.}, Particle data group, Chin. Phys. C, {\bf 38}, 090001.

\bibitem{William}
W.~R. Leo, Techniques for nuclear and particle physics experiments, 2nd
  Edition, Springer-Verlag, Berlin Heidelberg New York, 1994.

\bibitem{Jackson}
J.~D. Jackson, Classical Electrodynamics, 3rd Edition, John Wiley \&. Sons, New
  York, 1998.

\bibitem{moneta2005developments}
L.~Moneta, M.~Winkler, A.~Zsenei, P.~Mato-Vila, M.~Hatlo, F.~James,
  Developments of mathematical software libraries for the lhc experiments, IEEE
  Transactions on Nuclear Science 52 (2005) 2818--2822.

\bibitem{minuit}
C.~Group, Minuit users guide, Program Library D506, CERN.

\bibitem{suter2012musrfit}
A.~Suter, B.~Wojek, Musrfit: a free platform-independent framework for $\mu$sr
  data analysis, Physics Procedia 30 (2012) 69--73.

\bibitem{youanc2011muon}
A.~Youanc, P.~D. de~R{\'e}otier, Muon spin rotation, relaxation and resonance,
  Oxford University Press, Oxford, 2011.

\end{thebibliography}

\end{document}